\documentstyle[preprint,aps,floats,epsf]{revtex}
\def\ra{\rightarrow}
\begin{document}
\tightenlines
\preprint{\vbox{\baselineskip=14pt
   \hbox{\bf MADPH-99-1127}
   \hbox{\bf UH-511-931-99}
   \hbox{\bf hep-ph/9907421}
   \hbox{July 1999}
  }}


\title{\vspace*{.5in}
Neutrino Decay and Atmospheric Neutrinos}

\author{\vspace{.25in}
V. Barger$^1$, J.G. Learned$^2$, P. Lipari$^3$, M. Lusignoli$^3$,
S. Pakvasa$^2$, and T.J. Weiler$^4$}
\address{$^1$Department of Physics, University of Wisconsin, Madison, WI
53706\\
$^2$Department of Physics, University of Hawaii, Honolulu, HI 96822\\
$^3$Dipartimento di Fisica, Universit\`a ``La Sapienza", and  I.N.F.N.,
Sezione di Roma, Roma, Italy\\
$^4$Department of Physics and Astronomy, Vanderbilt University, Nashville, TN
37235}

\maketitle

\thispagestyle{empty}

\begin{abstract}
We reconsider neutrino decay as an explanation for atmospheric neutrino
observations.  We show that if the mass-difference relevant to the two mixed
states $\nu_\mu$ and $\nu_\tau$ is very small $(< 10^{-4}\rm\, eV^2)$, then a
very good fit to the observations
can be obtained with decay of a component of $\nu_\mu$ to a sterile neutrino
and a Majoron.
We discuss how the K2K and MINOS long-baseline experiments can distinguish the
decay and oscillation scenarios.
\end{abstract}


\newpage

Recently a neutrino decay explanation for the atmospheric neutrino
observations of Super-Kamiokande was proposed\cite{barger}.  To recapitulate
briefly, in presence of a decaying neutrino state, the survival
probability of $\nu_\mu$ is given by
\begin{eqnarray}
P(\nu_\mu\to \nu_\mu) = \sin^4 \theta + \cos^4 \theta e^{- \alpha L/E} + 2
\sin^2\theta \cos^2 \theta e^{-\alpha L/2E} \cos \left (
\frac{\delta m_{23}^2 L}{2E}\right )
\end{eqnarray}
where we have considered only two state mixing for simplicity:
$\delta m_{23}^2 =  m^2_2 - m^2_3,  \ \alpha = m_2 /\tau_2$  and
$\nu_\mu = \cos \theta \nu_2 + \sin \theta \nu_3$.  If the decay of
$\nu_2$ is $\nu_2 \ra \bar{\nu}_3 + J$, where J is a massless scalar,
then the $\delta m^2$
in the decay is the same as in Eq.~(1) and it can be shown that
this $\delta m^2$ has to be larger than 0.73~eV$^2$ to satisfy constraints from
$K\to\mu + \rm neutrals$ decay\cite{barger1}.  Then the
oscillating term in Eq.~(1) averages to zero and $P(\nu_\mu\to \nu_\mu)$
simplifies to
\begin{equation}
P(\nu_\mu\to \nu_\mu) = \sin^4 \theta \ + \cos^4 \theta e^{- \alpha L/E}
\end{equation}
This decay scenario was analyzed in Ref.~\cite{barger} and fit to the $L/E_\nu$
dependence and the
asymmetry of the contained events in Super-Kamiokande\cite{fuku}.  A
satisfactory
fit with $\cos^2 \theta \sim 0.87$ and $\alpha \sim 1~{\rm GeV}/D_E$ (where
$D_E$ is the diameter of the earth) was
found.  However, it has since been shown that the fit to the higher energy
 events in Super-K (especially the upcoming muons) is quite poor\cite{lipari},
 \cite{fogli}.

Another possibility, mentioned in Ref.~\cite{barger}, is that the decay of
$\nu_2$
is into a state with which it does not mix.
For example, the three weak coupling states $\nu_\mu, \nu_\tau, \nu_s$ (where
$\nu_s$ is a sterile neutrino) may be related to the mass eigenstates $\nu_2,
\nu_3, \nu_4$ by the approximate mixing
\begin{equation}
\left( \begin{array}{c} \nu_\mu\\ \nu_\tau\\ \nu_s \end{array} \right) =
\left( \begin{array}{ccc}  \cos\theta& \sin\theta& 0\\
                          -\sin\theta& \cos\theta& 0\\
                           0& 0& 1
\end{array} \right)
\left( \begin{array}{c} \nu_2\\ \nu_3\\ \nu_4 \end{array} \right)
\label{eq:mixing}
\end{equation}
and the decay is $\nu_2 \to \bar\nu_4 + J$. The electron neutrino,
which we identify with $\nu_1$, cannot mix very much with the other
three because of the more stringent bounds on its couplings\cite{barger1},
and thus our preferred solution for solar neutrinos would be
small angle matter oscillations.

Then the $\delta m_{23}^2$ in Eq.~(1) is not
related to the $\delta m_{24}^2$ in the decay, and can be very small,
say $ < 10^{-4}
\rm\, eV^2$ (to ensure that oscillations play no role in the atmospheric
neutrinos). In that case, the oscillating term is 1 and $P(\nu_\mu\to
\nu_\mu)$ becomes
\begin{equation}
P(\nu_\mu\to \nu_\mu) = (\sin^2 \theta + \cos^2 \theta e^{- \alpha L/2E})^2
\end{equation}
This is identical to Eq.~(13) in Ref.~\cite{barger}.  Here we consider this
decay
model and compare it to observations.

In order to compare the  predictions of this model with the standard
$\nu_\mu \leftrightarrow \nu_\tau$ oscillation model,
we have  calculated  with Monte Carlo methods   the  event  rates
for contained, semi-contained  and upward-going  (passing and stopping)
muons  in the Super-K  detector,  in the  absence of `new  physics', and
modifying the muon neutrino  flux   according
to the   decay or  oscillation  probabilities   discussed  above.
We have then  compared our  predictions  with the SuperK data
\cite{fuku},  calculating  a $\chi^2$  to
quantify the agreement   (or  disagreement)     between  data and
calculations.
In    performing  our  fit  (see Ref.~\cite{lipari}  for details)
we  do not take into account any  systematic  uncertainty, but we allow
the  absolute flux normalization to vary as  a free
parameter  $\beta$.

The  `no  new physics  model'     gives  a  very  poor  fit
to the data  with $\chi^2 = 281$ for
34  d.o.f.  (35  bins  and one   free parameter, $\beta$).
For the  standard $\nu_\mu \leftrightarrow \nu_\tau$ oscillation scenario
the  best fit    has  $\chi^2 = 33.3$   (32  d.o.f.)
and the values of the relevant parameters are
$\Delta m^2 = 3.2\times10^{-3}$~eV$^2$,
$\sin^2 2 \theta = 1$ and $\beta = 1.15$.
This  result is in  good  agreement  with the
detailed  fit  performed  by the SuperK collaboration \cite{fuku}
giving  us  confidence   that   our  simplified    treatment
of detector acceptances  and  systematic uncertainties is
reasonable.
The decay  model of Equations (3) and (4) above   gives an equally good fit
with a minimum $\chi^2 = 33.7$ (32 d.o.f.)
for the choice  of  parameters
\begin{equation}
\tau_\nu/m_\nu = 63\rm~km/GeV,
\ \cos^2 \theta = 0.30
\end{equation}
and normalization $\beta = 1.17$.

In Fig.~1  we  compare  the
best fits of the two  models  considered
(oscillations  and  decay)  with the
SuperK  data.
In the figure we show
(as  data points  with statistical error bars)
the ratios between the SuperK data and the Monte Carlo  predictions
calculated  in the  absence of oscillations or other
form of `new physics' beyond the standard model.
In the six  panels  we  show   separately  the  data
on $e$-like and  $\mu$-like events in the sub-GeV and multi-GeV
samples,   and  on   stopping and passing
upward-going muon events.
The   solid (dashed) histograms
correspond to  the  best fits for the decay  model
($\nu_\mu \leftrightarrow \nu_\tau$ oscillations).
One  can  see  that the  best fits  of the two  models
are  of comparable  quality.
The reason  for the similarity  of the results  obtained
in  the two  models  can be understood by looking  at
Fig.~2, where  we show
the survival probability $P(\nu_\mu \to \nu_\mu)$
of muon neutrinos   as  a  function
of $L/E_\nu$ for  the  two  models   using the
best  fit  parameters.
In the case  of the neutrino  decay model   (thick  curve)
the probability   $P(\nu_\mu \to \nu_\mu)$
monotonically  decreases   from    unity  to  an  asymptotic  value
$\sin^4 \theta \simeq  0.49$.
In the case of  oscillations the  probability  has  a sinusoidal
behaviour  in $L/E_\nu$.  The  two  functional    forms
seem     very different;  however,  taking  into  account  the
resolution in $L/E_\nu$,  the  two  forms
are  hardly  distinguishable.
In fact, in the    large $L/E_\nu$    region, the oscillations
are  averaged  out  and the survival  probability there
can  be  well  approximated  with 0.5  (for  maximal  mixing).
In  the region  of  small  $L/E_\nu$  both probabilities  approach
unity.
In the region $L/E_\nu$ around  400~km/GeV, where  the  probability for the
neutrino oscillation model  has the first  minimum,
the  two  curves are  most  easily  distinguishable, at least in
principle.

\medskip\noindent{\bf Decay Model}

There are two decay possibilities that can be considered: (a)~$\nu_2$ decays to
$\bar\nu_4$  which is dominantly $\nu_s$ with $\nu_2$ and $\nu_3$ mixtures of
$\nu_\mu$ and $\nu_\tau$, as in Eq.~(\ref{eq:mixing}), and
(b)~$\nu_2$ decays into $\bar\nu_4$ which is dominantly $\bar\nu_\tau$ and
$\nu_2$
and
$\nu_3$ are mixtures of $\nu_\mu$ and $\nu_s$.
In both cases the decay interaction has to be of the form
\begin{equation}
{\cal{L}}_{int} = g_{24} \ \overline{\nu_{4_{L}}^c} \ \nu_{2_{L}} J + h.c.
\end{equation}
where $J$ is a Majoron field that is dominantly iso-singlet (this avoids any
conflict with the invisible width of the $Z$).  Viable
models for both the above cases can be constructed \cite{valle,joshipura}.
However, case (b) needs additional iso-triplet light scalars which cause
potential problems with Big Bang Nucleosynthesis (BBN), and there is some
preliminary evidence from SuperK against $\nu_\mu$--$\nu_s$ mixing
\cite{kajita}. Hence we
only consider case (a), i.e.\ $\nu_2\to\bar\nu_4 + J$ with $\nu_4\approx
\nu_s$, as implicit in Eq.~(\ref{eq:mixing}).
With this interaction, the $\nu_2$ rest-lifetime is given by
\begin{equation}
\tau_2 = \frac
{16 \pi}{g^2} \cdot \
\frac{m_2}{\delta m^2 (1 + x)^2},
\end{equation}
where $\delta m^2 = m^2_2 - m^2_4$ and $x=m_4/m_2 \   (0 < x <1)$.
{}From the value of $\alpha^{-1} = \tau_2/m_2 = 63$~km/GeV found in the fit
and for $x=0$, we
have
\begin{equation}
g^2 \delta m^2 \simeq 0.16\rm\, eV^2
\end{equation}
Combining this with the bound on $g^2$ from $K \rightarrow \mu$ decays of $g^2
< 2.4\times10^{-4}$ \cite{barger1} we have
\begin{equation}
\delta m^2 > 650 \rm\ eV^2 \,.
\end{equation}
Even with a generous interpretation of the uncertainties in the fit,
this $\delta m^2$ implies a minimum mass difference in the range of about
25~eV.
Then $\nu_2$ and $\nu_3$ are nearly degenerate with masses
$\agt {\cal O}$(25~eV) and $\nu_4$ is relatively light. We assume
that
a similar coupling of $\nu_3$ to $\nu_4$ and J is somewhat weaker
leading to a significantly longer lifetime for $\nu_3$, and the instability of
$\nu_3$ is irrelevant for the analysis of the atmospheric neutrino
data.

For the atmospheric neutrinos in SuperK, two kinds of tests have been proposed
to distinguish between $\nu_\mu$--$\nu_\tau$ oscillations and
$\nu_\mu$--$\nu_s$ oscillations. One is based on the fact that matter effects
are present for $\nu_\mu$--$\nu_s$ oscillations\cite{bdppw} but are nearly absent for
$\nu_\mu$--$\nu_\tau$ oscillations\cite{panta} leading to differences in the zenith angle 
distributions  due to
matter effects on upgoing neutrinos \cite{lipari2}.
The other is the fact that the neutral current rate
will be affected in $\nu_\mu$--$\nu_s$ oscillations but not for
$\nu_\mu$--$\nu_\tau$ oscillations as can be measured in  events
with single $\pi^0$'s \cite{smirnov}. In these tests our decay scenario will
behave
as a hybrid in that there is no matter effect but there is some effect in
neutral current rates.

\medskip
\noindent{\bf Long-Baseline Experiments}

The survival probability of $\nu_\mu$ as a function of $L/E$ is given in
Eq.~(1). The conversion probability into $\nu_\tau$ is given by
\begin{equation}
P(\nu_\mu\to\nu_\tau) = \sin^2\theta \cos^2\theta (1-e^{-\alpha L/2E})^2 \,.
\end{equation}
This result differs from $1-P(\nu_\mu\to\nu_\mu)$ and hence is different from
$\nu_\mu$--$\nu_\tau$ oscillations. Furthermore, $P(\nu_\mu\to\nu_\mu)
+ P (\nu_\mu\to \nu_\tau)$ is
not 1 but is given by
\begin{equation}
P (\nu_\mu\to\nu_\mu) + P(\nu_\mu\to\nu_\tau) = 1 - \cos^2\theta (1 -
e^{-\alpha L/E})
\end{equation}
and determines the amount by which the predicted neutral-current rates are
affected compared to the no oscillations (or the $\nu_\mu$--$\nu_\tau$
oscillations) case.
In Fig.~3 we give the results for $P(\nu_\mu\to\nu_\mu)$,
$P(\nu_\mu\to\nu_\tau)$ and $P(\nu_\mu\to\nu_\mu) +
P(\nu_\mu\to\nu_\tau)$ for the decay model and compare them to the
$\nu_\mu$--$\nu_\tau$ oscillations, for both the K2K\cite{who} and
MINOS\cite{minos} (or the corresponding European project\cite{NGS})
long-baseline experiments, with the oscillation and decay parameters as
determined in the fits above.

The K2K experiment, already underway, has a low energy beam $E_\nu
\approx 1\mbox{--}2$~GeV and a baseline $L=250$~km.  The MINOS experiment will have
3 different beams, with average energies $E_\nu = 3,$ 6 and 12 GeV and a
baseline $L=732$~km.  The approximate $L/E_\nu$ ranges are thus 125--250~km/GeV for
K2K and 50--250~km/GeV for MINOS.  The comparisons in Figure 3 show that the
energy dependence of $\nu_\mu$ survival probability and the neutral
current rate can both distinguish between the decay and the oscillation
models.  MINOS and the European project may also have $\tau$ detection
capabilities that would allow additional tests.

\medskip
\noindent{\bf Big Bang Nucleosynthesis}

The decay of $\nu_2$ is sufficiently fast that all the neutrinos ($\nu_e,
\nu_\mu, \nu_\tau, \nu_s$) and the Majoron may be expected to equilibrate in
the early universe before the primordial neutrinos decouple. When they achieve
thermal equilibrium each Majorana neutrino contributes $N_\nu = 1$ and the
Majoron contributes $N_\nu = 4/7$ \cite{kolb}, giving and effective number of
light
neutrinos $N_\nu = 4{4\over7}$ at the time of Big Bang Nucleosynthesis. From
the observed primordial abundances of $^4$He and $^6$Li, upper limits on
$N_\nu$ are inferred, but these depend on which data are
used\cite{olive,lisi,burles}. Conservatively, the upper limit to $N_\nu$ could
extend up to 5.3 (or even to 6 if $^7$Li is depleted in halo stars\cite{olive}).

\medskip
\noindent{\bf Cosmic Neutrino Fluxes}

Since we expect both $\nu_2$ and $\nu_3$ to decay, neutrino beams
from distant sources (such as Supernovae, active galactic nuclei and
gamma-ray bursters) should contain only $\nu_e$ and $\bar\nu_e$
but no $\nu_\mu$, $\bar\nu_\mu$, $\nu_\tau$ and $\bar\nu_\tau$.
This is a very strong prediction of our decay scenario.

\medskip
\noindent{\bf Reactor and Accelerator Limits}

The $\nu_e$ is essentially decoupled  from the decay state $\nu_2$ so the null
observations from the CHOOZ reactor are satisfied\cite{chooz}. The mixings of $\nu_\mu$ and
$\nu_\tau$ with $\nu_s$ and $\nu_e$ are very small, so there is no conflict
with stringent accelerator limits on flavor oscillations with large $\delta
m^2$~\cite{zuber}.

\medskip
\noindent{\bf Conclusions}

 In summary, we have shown that neutrino decay remains a viable
 alternative to neutrino oscillations as an explanation of the atmospheric
 neutrino anomaly. The model consists of two nearly degenerate mass
 eigenstates $\nu_2$, $\nu_3$ with mass separation $\agt {\cal O}
(25$~eV) from
 another nearly degenerate pair $\nu_1$, $\nu_4$.
The $\nu_\mu$ and $\nu_\tau$ flavors are approximately composed of
$\nu_2$ and $\nu_3$, with a mixing angle $\theta_{23} \simeq 57^\circ$.
The state $\nu_2$ is unstable, decaying to $\bar{\nu}_{4}$ and a Majoron
with a lifetime $\tau_2 \sim 10^{-12}$ sec.  The electron neutrino
$\nu_e$ and a sterile neutrino $\nu_s$ have negligible mixing with $\nu_\mu,
\nu_\tau$ and are approximate mass eigenstates ($\nu_e \approx \nu_1,
\nu_s \approx \nu_4)$, with a small mixing angle $\theta_{14}$ and a
$\delta m_{41}^2 \approx10^{-5}\rm\,eV^2$ to explain the solar
neutrino anomaly.
The states $\nu_3$ and $\nu_4$ are also unstable, but with $\nu_3$ lifetime
somewhat longer and $\nu_4$ lifetime much longer than the
$\nu_2$ lifetime.
This decay
scenario is difficult to distinguish from oscillations because of the
smearing in both L and $E_\nu$ in atmospheric neutrino events.  However,
long-baseline experiments, where $L$ is fixed, should be able to establish
whether the dependence of $L/E_\nu$ is exponential or sinusoidal. In
our scenario only $\nu_1$ is stable.  Thus, neutrinos of supernovae
or of extra galactic origin would be almost entirely $\nu_e$.
The contribution of the electron neutrinos and the Majorons to the cosmological
energy density $\Omega$ is negligible and  not relevant for large
scale structure formation.

\medskip
\noindent{\bf Acknowledgements}

We would like to thank A. Joshipura for useful
discussions. This work is supported in part by the U.S. Department of Energy.

\newpage

\begin{figure}
\centering\leavevmode
\epsfxsize=3.5in\epsffile{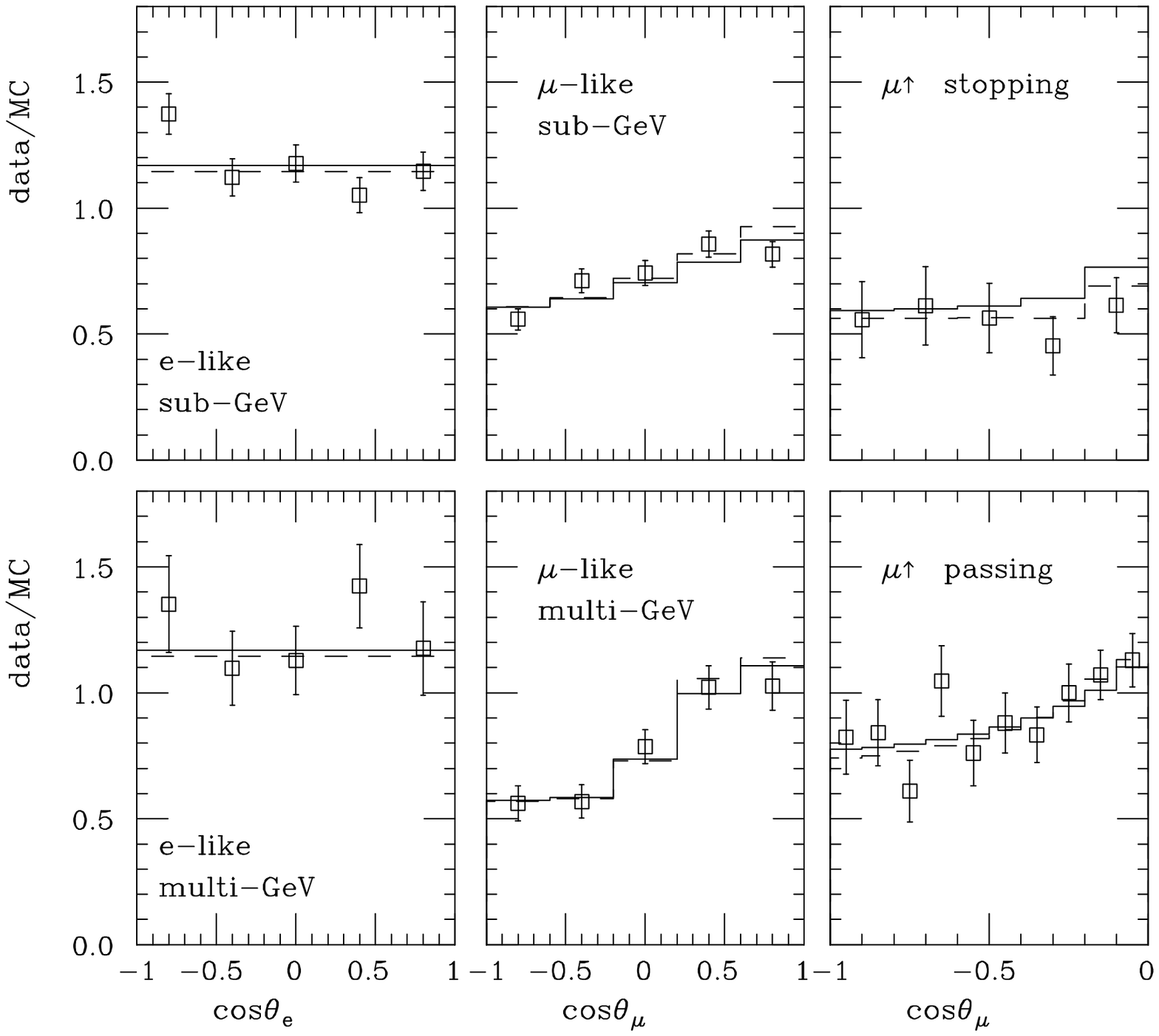}

\caption[]{Comparison of decay model (solid histograms) and
$\nu_\mu$--$\nu_\tau$ oscillation model (dashed histograms) with SuperK data
from Ref.~\cite{fuku}.}
\end{figure}

\begin{figure}
\centering\leavevmode
\epsfxsize=6.25in\epsffile{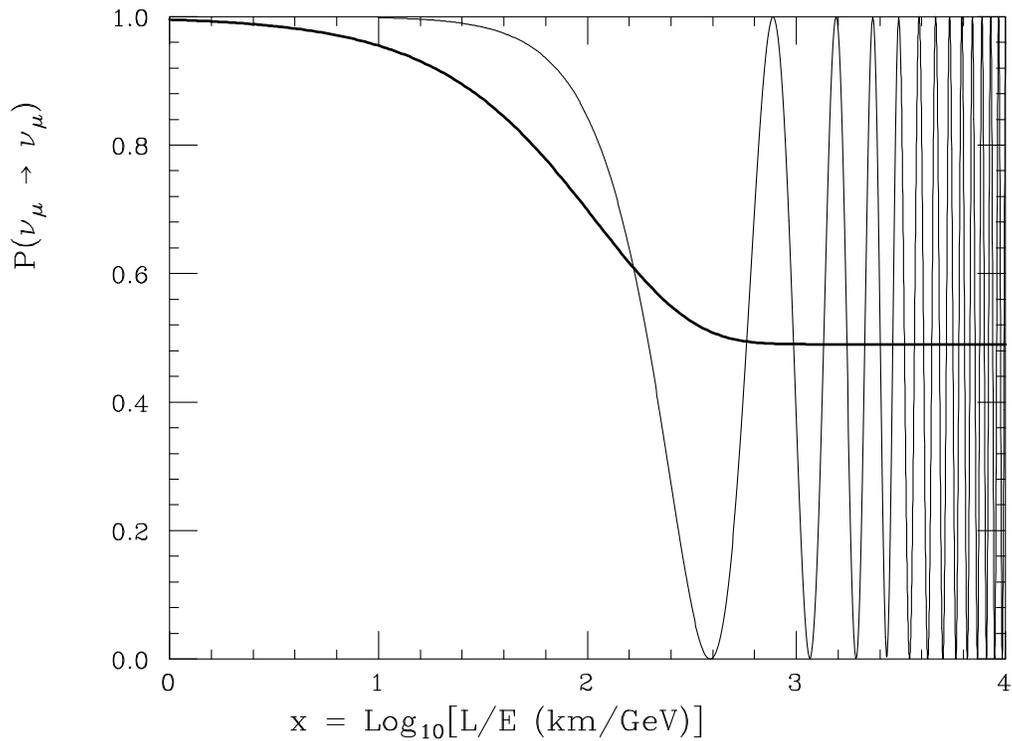}

\vspace{.5in}

\caption[]{Survival probabiliity for $\nu_\mu$ versus $\log_{10}(L/E)$ for the
decay model (heavy solid curve) and $\nu_\mu$ oscillation model (thin curve).}
\end{figure}

\begin{figure}
\centering\leavevmode
\epsfxsize=5.5in\epsffile{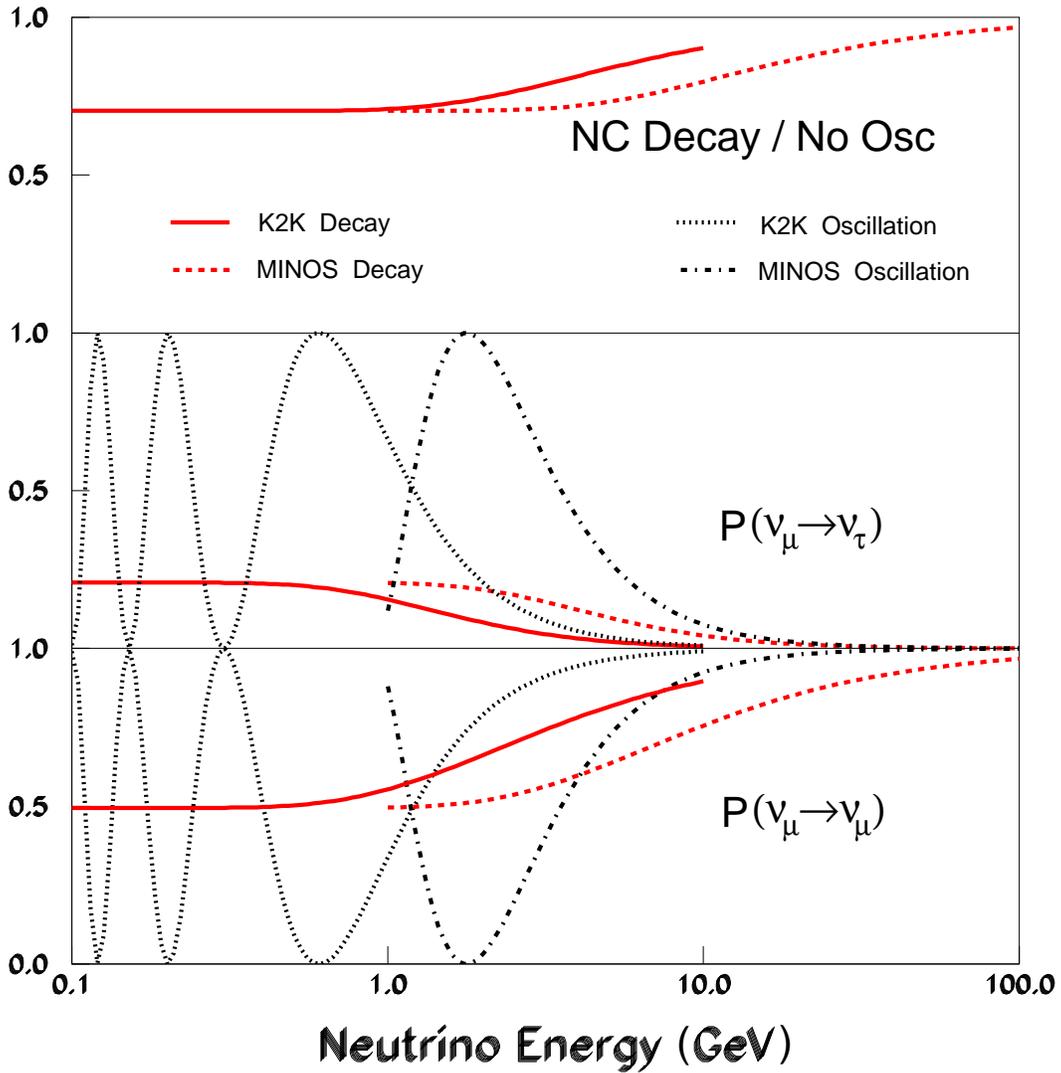}
\vspace{.5in}

\caption[]{Long-baseline expectations for the K2K and MINOS long-baseline
experiments from the
decay model and the $\nu_\mu$--$\nu_\tau$ oscillation model. The upper panel
gives the
neutral current predictions compared to no oscillations (or
$\nu_\mu$--$\nu_\tau$ oscillations).}
\end{figure}

\end{document}